\documentclass[aps,11pt,amsmath,amssymb]{revtex4}
\usepackage{dcolumn}% Align table columns on decimal point
\usepackage{bm}
\begin{document}
%\preprint{gr-qc/yymmddd}
\title{Hawking radiation as tunneling through the quantum horizon}
\author{Michele Arzano}
\email{arzano@physics.unc.edu}
\affiliation{Institute of Field Physics\\Department of Physics and Astronomy\\
University of North Carolina\\
Chapel Hill, North Carolina 27599-3255, USA}
\author{A.J.M. Medved}
\email{joey.medved@mcs.vuw.ac.nz}
\affiliation{School of Mathematics,
Statistics and Computer Science \\
Victoria University of Wellington \\
PO Box 600, Wellington, New Zealand }
\author{Elias C. Vagenas}
\email{evagenas@phys.uoa.gr}
\affiliation{Nuclear and Particle Physics Section \\
Physics Department\\
University of Athens\\
GR-15771, Athens, Greece}

\begin{abstract}
\begin{center}
{\bf Abstract}
\end{center}
Planck-scale corrections to the black-hole radiation spectrum in the Parikh-Wilczek tunneling framework are calculated.
The corrective terms arise from modifications in the expression of the surface gravity in terms of the mass-energy of the black hole-emitted particle system.
The form of the new spectrum is discussed together with the possible consequences for the fate of black holes in the late stages of evaporation.
\end{abstract}
\maketitle
%\vspace{-.3 in}
\section{Introduction}
It is widely believed that a final word on the so-called ``black hole information paradox" would
provide a key ingredient in the search for a yet to be formulated theory of Quantum
Gravity.  The paradox relies on the validity of the ``no hair" theorem \cite{Bekenstein:1971hc} and the fact that Hawking radiation exhibits a thermal spectrum, i.e. it is characterized by just one parameter: the black hole temperature  \cite{Hawking:1974sw}. Information carried by a physical system falling toward a
black-hole singularity then has no way to be recovered. This would allow, for example,
a pure quantum state to evolve into a mixed one, thus violating the unitarity of
evolution in Quantum Mechanics \cite{Hawking:1976ra}. The answer to the question of whether or not this violation occurs will deeply characterize the way in which Quantum Mechanics and General
Relativity would emerge as limits of the Quantum Gravity theory in regimes far from
the Planck scale.\\  Many proposals have been made in the attempt to save the basic principles of Quantum Mechanics in the presence of black holes (for reviews see \cite{Preskill:1992tc, Page:1993up, Danielsson:1993um, Russo:2005aw}).
Among these, the idea that information lost behind the horizon might re-emerge
via the Hawking radiation seems particularly interesting.  In fact in the late
stages of evaporation the usual picture for the emission process will lose its validity
and effects due to gravitational back-reaction should be taken into account.
These ``self-gravitation"
effects were studied in \cite{Kraus:1994by, Keski-Vakkuri:1996xp}
and later in \cite{Parikh:1999mf, Parikh:2004ih, Parikh:2004rh}
(also see, e.g., \cite{Padmanabhan:1998,Vagenas:2001,Medved:2001}).
In particular Parikh and Wilczek \cite{Parikh:1999mf}
showed how the inclusion of back-reaction effects, which ensures energy conservation during the emission of a particle via tunneling through the horizon, leads to non-thermal corrections to the black-hole radiation
spectrum.  However the form of the correction the authors find is such that no statistical correlation between quanta emitted with different energies appears \cite{Parikh:2004rh}.
\\  In this letter we proceed a step further and modify the Parikh-Wilczek
tunneling picture using a typical Planck-scale modification for the near-horizon
emission process.  The type of modification we consider is directly related to the logarithmic corrective term appearing
 in the Bekenstein-Hawking entropy-area relation from the direct count of black hole micro-states in String Theory and Loop Quantum Gravity \cite{Strominger:1996sh, Solodukhin:1997yy, Rovelli:1996dv, Ashtekar:1997yu,Kaul:2000kf}.
Moreover, the logarithmic correction has been substantiated in a variety
of different contexts (see \cite{Page:2004} for a list of relevant citations).
For instance,
in \cite{Amelino-Camelia:2004xx} (also see \cite{Vagenas:2004})
it is showed how this corrective term can be also derived from a general form of deformed energy-momentum dispersion relation which emerges in various models of Planck scale departures from Lorentz symmetry. \\
It might have been hoped that,
when back-reaction effects are combined with Planck-scale corrections,
 one would
 observe the appearance of correlations between the probability of
emission of different modes in the black-hole
radiation spectrum.
%THIS WOULD THEN MEAN THAT
 %energy conservation
%in this case
%holds in a strong sense.
Unfortunately, even with the inclusion of the Quantum Gravity corrections,
no such correlations are conclusively in evidence. One does find, however,
that (given a certain constraint to be discussed below)
%In fact, due to the introduction of Quantum Gravity corrections,
the probability of emission of a particle approaches zero when its energy
becomes of the order of the mass of the emitting black hole.
This is  a novel feature of our  analysis that was not present in the
``semiclassical" treatment of
\cite{Parikh:1999mf, Parikh:2004ih, Parikh:2004rh}.
\\ In the next sections we briefly review how the motion of a
classical spherical shell in a Schwarzschild geometry is modified in the presence of gravitational
back-reaction and the way this is used to compute corrections to the black-hole emission
spectrum. In section III we review how these effects were described by Parikh and Wilczek
in terms of quantum tunneling through the black hole horizon.
In section IV we discuss a possible role of the Quantum Gravity correction to the entropy-area law for the non-thermal spectrum and its consequences for the escape of information via Hawking radiation.
In section  V  we
%then
propose a modified version of the tunneling picture in which Planck-scale corrections are introduced and obtain a black hole emission spectrum in which both back-reaction and Quantum Gravity effects are present.
We then illustrate the formalism with a specific example (section VI),
which is followed by some concluding remarks (section VII).

\section{Motion of a self-gravitating spherical shell in a Schwarzschild geometry}
We summarize here the results of \cite{Kraus:1994by} where the corrections
to the geodesic motion of a spherical shell due to self-gravitation
in a Schwarzschild geometry were calculated. We start by writing the metric for
a general spherically symmetric system in ADM form
\begin{equation}
ds^2=-N_t(t,r)^2dt^2+L(t,r)^2[dr+N_r(t,r)dt]^2+R(t,r)^2d\Omega^2\,  .
\label{metricADM}
\end{equation}
The action for the black hole plus the emitted shell system is
\begin{equation}
S=\frac{1}{16\pi}\int d^4x\sqrt{-g}\,\mathcal{R}-m\int dt\sqrt{(\hat{N}^t)^2-(\hat{L}^r)^2(\dot{\hat{r}}+\hat{N}^r)^2}+\text{boundary terms}\, ,
\end{equation}
where $\hat{r}$ is the shell radius and the other quantities under `` $\hat{}$ " are evaluated at the shell through
$\hat{g}_{\mu\nu}=g_{\mu\nu}(\hat{t},\hat{r})$.
The above action can be written in Hamiltonian form where all the canonically conjugate momenta appear.
Since the system has only one effective degree of freedom the idea is to solve the constraints of the theory in order to eliminate the dependence of the action from all the momenta but the one conjugate to the shell radius.  This remaining degree of freedom can be expressed in terms of the total mass/energy of the system and it is obviously related to the position of the shell.  In the approach followed in \cite{Kraus:1994by},
the total (ADM) mass is allowed to vary with time while the mass of the hole is kept fixed. This time dependence accommodates the dynamics of the system and allows energy conservation to hold at any time of the process.\\
Once the constraints are solved and the expression for the conjugate momenta are substituted into the action
one integrates over the gravitational degrees of freedom to obtain an effective action. Furthermore one specializes to the case of a massless
particle $(m=0)$ and fixes the gauge appropriately ($L=1$ $R=r$).  This choice of the gauge corresponds to a
particular set of coordinates for the line element (Painleve' coordinates) which is particularly useful to study across horizon phenomena being non-singular at the horizon and having Euclidean constant time slices (for more details see \cite{Kraus:1994fh}).
The effective action for a massless
gravitating spherical shell is then
\begin{equation}
S=\int dt \left(p_c\dot{\hat{r}}-M_+\right)\, ,
\end{equation}
where $p_c$ is the momentum canonically conjugate to $\hat{r}$, the radial position of the shell, and
$M_+$ is the total mass of the shell-hole system which plays the role of the Hamiltonian.
In terms of the black hole mass $M$ and the shell energy $E$ we have $M_+=M+E$.
Details of the lengthy derivation can be found in \cite{Kraus:1994by}.
The trajectories which extremize this action are the null geodesics of the metric
\begin{equation}
ds^2=-[N_t(r; M+E)dt]^2+[dr+N_r(r; M+E)dt]^2+r^2d\Omega^2\,  ,
\label{linel}
\end{equation}
for which
\begin{equation}
\frac{dr}{dt}= N_t(r; M+E)-N_r(r; M+E)\, .
\label{geodesic1}
\end{equation}
\section{The KKW model in a nutshell}
In \cite{Kraus:1994by} and \cite{Keski-Vakkuri:1996xp}, Keski-Vakkuri, Kraus and Wilczek
showed how corrections of the type considered above can affect the emission spectrum of a black hole.
They consider the Bogoliubov coefficients, $\alpha_{k k'}$ and $\beta_{k k'}$, connecting the
positive and negative frequency modes of an asymptotic observer and one freely
falling through the horizon. The (semiclassical) WKB approximation is then used
to express the mode solutions for the observer crossing the horizon.  Such an
approximation is valid since the black hole emission is dominated by modes that have very small wavelengths
close to the horizon and undergo a large red-shift when propagating away from it.  Once this approximation is
taken into account one finds \cite{Keski-Vakkuri:1996xp}
\begin{equation}
|\alpha_{k k'}|=e^{-Im\int_{r_+(0)}^{r_f} p_+(r)dr}\,\,\,\, ;\,\,\,\,\,|\beta_{k k'}|=e^{-Im\int_{r_-(0)}^{r_f}p_-(r)dr}\, ,
\end{equation}
where $r_{\pm}$ and $p_{\pm}$ are the trajectories and momenta of positive and negative energy modes propagating
in and out of the horizon and $r_f$ is located outside the horizon.  From Hamilton's equation $\dot{r}=\frac{\partial H}{\partial p}=\frac{\partial E}{\partial p}$, with the Hamiltonian given by
$H=M+E$, one can express $p_{\pm}$ using (\ref{geodesic1})
\begin{equation}
p_{\pm}(r)=\int_{0}^{\pm\omega_k}\frac{dE}{N_t(r; M+E)-N_r(r; M+E)}\,\,  .
\end{equation}
It can be shown  \cite{Keski-Vakkuri:1996xp} that $\alpha_{k k'}=1$.  For  $\beta_{k k'}$ one instead finds
\begin{equation}
\label{bocoff}
\mathrm{Im}\int_{r_-(0)}^{r_f}p_-(r)dr=
-\pi\int_0^{-\omega_k}\frac{dE}{\kappa(M+E)}\, .
\end{equation}
In order to calculate the amplitude for particle production in the KKW model, it is assumed that the emission
in the low energy regime
is uncorrelated so that the amplitude in terms of $\left|\beta_{k k'}/\alpha_{k k'}\right|^2$ reads
\begin{equation}
\rho(\omega_k)=\frac{\left|\beta_{k k'}/\alpha_{k k'}\right|^2}{1-\left|\beta_{k k'}/\alpha_{k k'}\right|^2}\, .
\end{equation}
Instead when $\omega_k$ is comparable with the mass of the black hole and at most one quantum can be emitted
\begin{equation}
\rho(\omega_k)\simeq \left|\frac{\beta_{k k'}}{\alpha_{k k'}}\right|^2\, .
\end{equation}
Using the first law of black-hole thermodynamics $dM=\frac{\kappa(M)}{2\pi}dS$, one can evaluate (\ref{bocoff}).
For the first case
when $\omega_k$ is small compared to $M$, we obtain the usual emission amplitude governed by the Hawking temperature.
For large $\omega_k$ instead one has
\begin{equation}
\rho(\omega_k)\simeq \exp\left[S(M-\omega_k)-S(M)\right]\, .
\end{equation}
Substituting the standard Bekenstein-Hawking expression for the black hole entropy in the previous equation leads to
a non-thermal correction,
quadratic in $\omega_k$, to the typical Boltzmann factor of the emission probability.
\section{Quantum tunneling and non-thermal spectrum}
The results of \cite{Kraus:1994by} and \cite{Keski-Vakkuri:1996xp} can be recast in an elegant and simple form
if one describes the emission of a particle as a tunneling process \cite{Parikh:1999mf, Parikh:2004ih}.
This is done by considering the geometrical optics limit\footnote{This limit is valid for the same reason
that allowed us to use the WKB approximation in the analysis of the previous section, i.e. the fact that the emitted wave packets are arbitrarily blue-shifted close to the horizon.}
so that one can treat the wave-packets near the horizon as effective
particles.  The emission of each of these particles is seen as a tunneling through a barrier set
by the energy of the particles itself.  This simple argument makes it possible to avoid the machinery of Bogoliubov
coefficients and also shows how energy conservation is naturally preserved during the emission process \cite{Parikh:2004rh}.\\
The authors consider now an explicit expression for the line element (\ref{linel}) obtained from the expressions of $N_t$ and $N_r$
given by the constraint equations \cite{Kraus:1994by}
\begin{equation}
N_t=\pm1\,\,\,;\,\,\,  N_r=\pm\sqrt{\frac{2M_+}{r}}\,\,.
\end{equation}
In the model proposed in \cite{Parikh:1999mf}
the total mass of the system is kept fixed while the hole mass is allowed to vary. This means that
the mass parameter $M_+$ is now $M_+=M-E$.
One then has the following expression for a spherical shell moving along a radial null geodesic
\begin{equation}
\dot{r}=\pm1-\sqrt{\frac{2(M-E)}{r}}\, .
\label{geodesic2}
\end{equation}
In the WKB approximation the tunneling probability is a function of the imaginary part of the particle's action
\begin{equation}
\label{tunnelampl}
\Gamma\sim e^{-2\,\mathrm{Im}\, S}\,\, .
\end{equation}
If we consider the emission of a spherical shell we have
\begin{equation}
\label{shellaction}
\mathrm{Im}\, S=\mathrm{Im}\int_{r_{in}}^{r_{fin}}p_r dr\, ,
\end{equation}
where $r_{in}$ and $r_{fin}$ are just inside and outside the barrier through which the particle is tunneling.
We can now see the key point:
the expression for $\mathrm{Im}\, S$ is the same as the one for the Bogoliubov coefficient $\beta_{kk'}$ used in the KKW analysis.  The same coefficient characterizes the emission probability in the field theoretical model.  The advantage of considering the particle-tunneling picture is that the correction we get now is present at all energy regimes even though it becomes dominant only in the high energy regime.
To calculate $\mathrm{Im}\, S$ we use once again Hamilton's equation, $\dot{r}=\frac{\partial H}{\partial p}$,
\begin{equation}
\mathrm{Im}\, S=\mathrm{Im}\int_{r_{in}}^{r_{fin}}p_r dr=
\mathrm{Im}\int_{r_{in}}^{r_{fin}}\int_M^{M-E}\frac{dH'}{\dot{r}} dr\, .
\end{equation}
The Hamiltonian is $H'=M-E'$, so the imaginary part of the action reads
\begin{equation}
\mathrm{Im}\, S=-\mathrm{Im}\int_{r_{in}}^{r_{fin}}\int_0^E\frac{dE'}{\dot{r}} dr\, .
\label{imsf}
\end{equation}
Using (\ref{geodesic2}) and integrating first over $r$ one easily obtains
\begin{equation}
\Gamma\sim \exp\left(-8\pi M E\left(1-\frac{E}{2M}\right)\right)\, ,
\label{proba}
\end{equation}
which, provided the usual Bekenstein-Hawking formula $S_{BH}=A/4=4\pi M^2$ is valid, corresponds to the KKW result
\begin{equation}
\label{Gammaf}
\Gamma\sim \exp\left[S_{BH}(M-E)-S_{BH}(M)\right]\, .
\end{equation}
If one integrates (\ref{imsf}) first over the energies it is easily seen that in order
to get (\ref{proba}) we must have $r_{in}=M$ and $r_{out}=M-E$.  So according to what one would expect from energy conservation, the tunneling barrier is set by the shrinking of the black hole horizon with a change
in the radius set by the energy of the emitted particle itself.\\
An interesting aspect to analyze is whether or not the non-thermal correction obtained leads to statistical correlations between the probabilities of emission of quanta with different energies.  This would allow for information to be encoded in the emitted radiation.
Consider for example the probability of emission of a quantum of energy $E=E_1+E_2$ and two quanta of energies $E_1$ and $E_2$.  The function
\begin{equation}
\chi(E_1+E_2;E_1,E_2)=\log(\Gamma(E_1+E_2))-\log(\Gamma(E_1)\Gamma(E_2))
\end{equation}
measures the statistical correlation between the two probabilities.  It
is zero when the probabilities are independent (or ``uncorrelated") as e.g. for a thermal emission spectrum like the one of a radiating black body.
Using (\ref{proba}) it is easy to verify  \cite{Parikh:2004rh} that for the non-thermal correction due to back-reaction effects
$\chi(E_1+E_2;E_1,E_2)=0$.  One concludes that back-reaction effects alone do not provide a straightforward way in which information can emerge from the horizon.  There might well be other processes that would allow the information of a pure quantum state to be recovered after its gravitational collapse but one would have to resort to other mechanisms \footnote {As proposed in  \cite{Parikh:2004rh} correlations might appear when back-reaction effects are considered in transient phases of the black hole emission.} .
\section{The Parikh-Wilczek tunneling picture revisited}
In this section we discuss a modification of the above argument that takes
into account the presence of Quantum Gravity-induced corrections.
%We first notice, in the first subsection, how a logarithmic corrective term
%in the entropy-area relation might allow information leak through Hawking
%radiation when back-reaction effects are taken into account.
%In this case one obtains again a non-thermal spectrum but with the novel
%feature that modes with different energies have now non-zero statistical
%correlation.
%The general argument presented in the first subsection applies to
%any Quantum Gravity scenario predicting a logarithmic
%corrective term in the entropy-area relation.  We suggest, therefore, that
%those features of the Quantum Gravity scenario that
%are responsible for the appearance of a logarithmic correction should also
%affect the the Parikh-Wilczek analysis, thereby leading to
%information leak.  We explicitly verify this conjecture in the
%case \cite{Amelino-Camelia:2004xx} of a logarithmic correction
%emerging from Planck-scale modifications of the energy-momentum
%dispersion relation. \\
%\subsection{Information leak and logarithmic correction}
Let us first notice that
the probability of emission of a shell with energy $E$, in the presence of
back-reaction effects, put in the form (\ref{Gammaf}) is highly suggestive.
It is what one would expect from a quantum mechanical
calculation of a transition rate where, up to a factor containing the square of the amplitude of the process,
\begin{equation}
\label{ampliG}
\Gamma\sim \frac{e^{S_{fin}}}{e^{S_{in}}}=\exp\left(\Delta S\right)\, .
\end{equation}
In other words the emission probability is proportional to a phase space factor which depends on the initial and final entropy of the system.  The entropy is directly related to the number of micro-states available to the system itself.\\
This observation calls for an immediate generalization.  Derivations of the black hole entropy-area relation by direct micro-state counting in String Theory and Loop Quantum Gravity  \cite{Strominger:1996sh, Solodukhin:1997yy, Rovelli:1996dv, Ashtekar:1997yu,Kaul:2000kf}
besides reproducing the familiar Bekenstein-Hawking linear relation give a leading order correction with a logarithmic \footnote{A similar logarithmic term has also emerged in various other treatments (see \cite{Page:2004}
for a list);  for instance,  the calculation of
one-loop quantum corrections to the entropy-area law in ordinary QFT
\cite{Fursaev:1994te}. }
dependence on the area \footnote{We now switch from $k=\hbar=c=G=1$ units of the previous sections to $k=\hbar=c=1$ to keep track of the Planck-scale suppressed terms.}
\begin{equation}
\label{logcorrs}
S_{QG}=\frac{A}{4L_p^2}+\alpha \ln \frac{A}{L_p^2}+O\left(\frac{L_p^2}{A}\right)\,\, .
\end{equation}
In the case of Loop Quantum Gravity $\alpha$ is a negative coefficient whose
exact value was once an object of debate (see e.g. \cite{Ghosh:2004rq})
but has since been rigorously fixed at $\alpha=-1/2$ \cite{Meissner:2004}.
In String Theory the sign of $\alpha$ depends on the number of field species appearing in the low energy approximation \cite{Solodukhin:1997yy}.\\
Now consider the emission of a particle of energy $E$ from the black hole.
One might expect that a derivation of the emission probability in a Quantum Gravity framework
in presence of back-reaction would lead to an expression analogous to (\ref{ampliG}) with the usual
Bekenstein-Hawking entropy $S_{BH}=\frac{A}{4L_p^2}$ replaced by (\ref{logcorrs}), i.e.
\begin{equation}
\Gamma\sim \exp{(S_{QG}(M-E)-S_{QG}(M))}\, .
\end{equation}
The previous expression written in explicit form reads
\begin{equation}
\Gamma(E)\sim \exp{(\Delta S_{QG})}=
\left(1-\frac{E}{M}\right)^{2\alpha}\exp \left(-8\pi GME\left(1-\frac{E}{2M}\right)\right)\, .
\label{prob}
\end{equation}
The exponential in this equation shows the same type of non-thermal deviation found in \cite{Parikh:1999mf}.
In this case, however, an
additional factor depending on the ratio of the energy of the emitted quantum and the mass of the black hole is present. \\
We would now like to know whether or not, in our case,
%A very interesting consequence of the above expression is that
the emission probabilities for different modes are statistically correlated \cite{Medved:2005vw}.
%In our case,
Using (\ref{prob}), we have, for a first emission of energy $E_1$,
\begin{equation}
\ln[\Gamma(E_1)]
%&=& S_q(M-E_1)-S_q(M) \label{07} \\ &=&
= -8\pi
GME_1\left(1-{E_1\over 2M}\right) \;+\; 2 \alpha
\ln\left[{M-E_1\over M}\right]\;.
%\nonumber
\end{equation}
Then  a second emission of energy $E_2$ gives us
\begin{equation}
\ln[\Gamma(E_2)]
%&=&  S_q(M-E_1-E_2)-S_q(M-E_1) \label{08} \\ &=&
= -8\pi
G(M-E_1)E_2\left(1-{E_2\over 2(M-E_1)}\right)
\;+\; 2 \alpha
\ln\left[{M-E_1-E_2\over M-E_1}\right]\;.
%\nonumber
\end{equation}
Alternatively, a single emission of the same total energy yields
\begin{equation}
\ln[\Gamma(E_1+E_2)]
%&=&  S_q(M-E_1-E_2)-S_q(M-E_1) \label{08} \\ &=&
= -8\pi
GM(E_1+E_2)\left(1-{E_1+E_2\over 2 M}\right)
\;+\; 2 \alpha
\ln\left[{M-E_1-E_2\over M}\right]\;.
%\nonumber
\end{equation}
It is now easily verified that the vanishing of the correlations, or
\begin{equation}
\chi(E_1+E_2;E_1,E_2)
=0 \; ,
%=-2\alpha\log\left(1+\frac{E_1E_2}{M(M-E_1-E_2)}\right)\,\, .
\end{equation}
is still in effect  at least to this logarithmic order. In fact, after just
a few iterations, one should readily  be convinced that this outcome
will persist up to any perturbative
 order of the quantum-corrected entropy [cf, equation (\ref{logcorrs})].
Hence, it does not appear that the inclusion of such correlations can
account for the mode correlations after all. \\
%\\
%{\bf (M. thinks the following should be omitted b/c it's true only
%if the QG spectrum exhibits correlations)}*****ON THE OTHER HAND,
%The correlation between the modes is always non zero in contrast to what
%is obtained considering only
%the non-thermal corrections coming from back-reaction effects.\\
%ONE MIGHT
%Now
%consider Shannon's formula, which defines the entropy associated with a
%certain emission spectrum
%\begin{equation}
%s=-\sum_{n}\Gamma(E_n)\log \Gamma(E_n)\, .
%\end{equation}
%One can easily see that the entropy associated with a spectrum of
%independent modes (like in the standard Hawking radiation or in the
%back reaction corrected picture) $s_{IN}$ is larger than the
%entropy $s_{QG}$ associated with a spectrum
%characterized by (\ref{prob}).  ONE MIGHT BE TEMPTED TO
%CONJECTURE THAT
%the difference $s_{IN}-s_{QG}$ measures
%the amount of information encoded in the correlations between different modes.
%IF THIS WERE THE CASE, IT
%This
%might provide a possible window through which the details of a
%pure quantum state fallen beyond the horizon could be read off
%during the black hole evaporation.*****
%ALTHOUGH SUCH IDEAS SEEM TO BE IN CONFLICT WITH OUR FINDINGS ABOVE,
On the other hand, it is interesting to note that the  emission
of three or more quanta could still induce a non-zero correlation
even for the tree-level calculation. By which we mean that,
for the sequential emission of (e.g.)
$E_1$, $E_2$ and $E_1+E_2$, then   $S(M-E_1)-S(M)+
S(M-E_1-E_2)-S(M-E_1) \neq S(M-2E_1-2E_2)-S(M-E_1-E_2)$.
And so  there  still appears to be  viability, on  some
level, for the notion that information leaks out in the tunneling process. \\
%Finally we
Let us also
 notice how the appearance of the Quantum Gravity suppression term
 in (\ref{prob}) can  cause $\Gamma(E)\rightarrow 0$
 when the energy of the emitted quantum approaches the mass of the black hole.
However, this suppression can only take place when $\alpha >0 $; whereas
a negative value of $\alpha$ will, conversely, cause the probability
to diverge as the same limit is approached!~\footnote{Let us, however,  point
out one possible loophole: higher-order corrections may conspire to
induce a suppression that is stronger than this divergence.} \\
Considering that Loop Quantum Gravity predicts a negative value for
$\alpha$, one might have preferred the above pattern to be  reversed.
But our outcome is actually quite reasonable, inasmuch as an evaporating
Schwarzschild black hole is (by virtue of
a negative heat capacity) highly unstable when in isolation.
This instability can, nevertheless, be resolved
 with the immersion of the black hole in a suitable heat bath.
In this event, the system will equilibrate  until the temperature
of the bath reaches that of the Hawking temperature; at which point,
stability ensues. Since a system in thermal equilibrium is most
suitably described by a canonical ensemble, it  now
becomes necessary to account for the canonical corrections
to the black hole entropy. It is therefore quite remarkable
and, at the same time, reassuring that the inclusion of these canonical
corrections appears to  ensure a strictly
non-negative value for $\alpha$ \cite{Medved:2004}.
To elaborate, the canonical ensemble
allows for thermal fluctuations in the horizon area,
which translates into  an increase in the black hole entropy.  A rigorous
calculation demonstrates that this increase is, at the very
least, $\frac{1}{2}\ln A$ (e.g.,
\cite{Kastrup:1997,Gour:2003,Majumdar:2003}) and
almost certainly larger once the fluctuations in the charge and spin are
properly accounted for \cite{Medved:2004}. (The point being that even a
classically
neutral and static black hole would still experience fluctuations
in these quantities.)  Hence, the canonical correction is
more than sufficient to compensate for the Loop
Quantum Gravity prediction of $\alpha=-1/2$ \cite{Meissner:2004}.
%This feature is crucial for the conservation of energy in the
% emission process.\\

\section{Tunneling in the presence of near-horizon Planck-scale effects}
We give now an explicit example of a modification of the Parikh-Wilczek derivation, in the presence of Planck-scale effects, which indeed gives rise to an emission spectrum of the type (\ref{prob}).
In \cite{Amelino-Camelia:2004xx} one of the authors (MA), Amelino-Camelia and Procaccini showed how in the context of Loop Quantum Gravity a logarithmic corrective term to the entropy-area law of the type present in (\ref{logcorrs}) can be related to a modification of the energy-momentum dispersion
relation for a massless particle propagating in flat space-time
\begin{equation}
\label{modisprel}
p^2\simeq \left(1+\eta L_p^2 E^2\right)E^2\,\, .
\end{equation}
with $\eta=\frac{2\pi}{3}\alpha$.\\
As we already observed in the previous sections, the black hole radiation spectrum seen from an observer at infinity is dominated by modes that propagate from ``near" the horizon where they have arbitrarily high frequencies and their wavelengths can easily go below the Planck length \cite{Jacobson:1991gr,Jacobson:1993hn}.  It turns out then that a key assumption in all the derivations of the Hawking radiation is that
the quantum state near the horizon looks, to a freely falling observer, like the Minkowski vacuum.  In other words Lorentz symmetry should hold up to extremely short scales or very large boosts.
It is plausible then that the motion of our particle tunneling through the horizon might be affected by Planck scale corrections of the type (\ref{modisprel}).  These type of modified dispersion relations
have, in fact, been proposed as low-energy Quantum Gravity effects, which deform
\cite{Amelino-Camelia:2000ge, Amelino-Camelia:2000mn} or break \cite{Amelino-Camelia:1997gz, Gambini:1998it}
Lorentz symmetry (see also \cite{Ng:2003jk}).\\
One would expect that an analysis analogous to the ones of the previous sections with opportune
modifications should lead to
a result of the form (\ref{Gammaf}) with $S_{BH}$ replaced by $S_{QG}$.  Here we provide an example of such an
analysis within the tunneling framework of Parikh and Wilczek.\\
Once again we consider the emission of a spherical shell and compute the tunneling amplitude (\ref{tunnelampl}) through
(\ref{shellaction})
\begin{equation}
\mathrm{Im}\, S=\mathrm{Im}\int_{r_{in}}^{r_{fin}}p_r dr=
\mathrm{Im}\int_{r_{in}}^{r_{fin}}\int_0^H\frac{dH'}{\dot{r}} dr=
-\mathrm{Im}\int_{r_{in}}^{r_{fin}}\int_0^{E}\frac{dE'}{\dot{r}} dr
\,\, .
\end{equation}
where we used the fact that for the Hamiltonian $H=M-E$.
Now we proceed to evaluate the integral without using an explicit form for
the null geodesic of the spherical shell in terms of its energy.
In fact, near the horizon, where our integral is being evaluated, one has
\begin{equation}
N_t(r; M)-N_r(r; M)\simeq (r-R)\,\kappa(M)+O((r-R)^2)
\end{equation}
where R is the Schwarzschild radius and $\kappa(M)$ is the horizon surface gravity.
Taking into account self-gravitation effects, $\dot{r}$ can be approximated by
\begin{equation}
\dot{r}\simeq (r-R)\,\kappa(M-E)+O((r-R)^2)\, .
\end{equation}
We can then write
\begin{equation}
\mathrm{Im}\, S=-\mathrm{Im}\int_{r_{in}}^{r_{fin}}\int_0^{E}\frac{dE'}{(r-R)\,\kappa(M-E')} dr
\,\, .
\end{equation}
Integrating over $r$, using the Feynman prescription\footnote{The pole is
moved in the lower half plane as in \cite{Parikh:1999mf}.} for the pole on the real axis $r=R$, we get
\begin{equation}
\mathrm{Im}\, S=-\pi\int_0^{E}\frac{dE'}{\kappa(M-E')}
\,\, .
\label{Ims}
\end{equation}
The surface gravity appearing in the above integral carries Quantum Gravity
corrections coming from Planck-scale modifications of near horizon physics related to (\ref{modisprel}).
As shown explicitly in \cite{Amelino-Camelia:2004xx}
these modifications are such that they reproduce via the first law of black hole thermodynamics, $dE'=dM'=\frac{\kappa(M)}{2\pi}dS$, the Quantum Gravity corrected entropy-area law (\ref{logcorrs}).
Using the first law, (\ref{Ims}) becomes
\begin{equation}
\mathrm{Im}\, S=-\frac{1}{2}\int_{S_{QG}(M)}^{S_{QG}(M-E)}dS=\frac{1}{2}[S_{QG}(M)-S_{QG}(M-E)]
\end{equation}
which leads to a probability of emission
\begin{equation}\label{prob2}
\Gamma(E)\sim \exp{(-2\mathrm{Im}S)}=
\left(1-\frac{E}{M}\right)^{2\alpha}\exp \left(-8\pi GME\left(1-\frac{E}{2M}\right)\right)\,
\end{equation}
which is analogous to (\ref{prob}).

\section{Conclusion}
We
%showed
have discussed
 how Quantum Gravity and back-reaction effects combined might provide a way
for information to be recovered from a black hole via Hawking radiation.
Although this matter still remains unsettled, we would argue
that our direction of study
%This
could have important implications for the fate of the unitarity of
quantum evolution in Quantum Gravity.  Our
%derivation
formal treatment
gives an idea of how near horizon physics provides an excellent
arena for studying the interplay of seemingly different aspects of Quantum Gravity as the number of microscopic degrees of freedom of a black hole and possible Planck-scale modifications of space-time symmetries.\\ \\
\begin{center}
{\bf Acknowledgments}
\end{center}
MA would like to thank Giovanni Amelino-Camelia for discussions and valuable
suggestions and Jack Ng for useful comments. MA
 also thanks the Department of Physics of the University of
Rome "La Sapienza" for hospitality during December 2004.
Research for AJMM is supported  by
the Marsden Fund (c/o the  New Zealand Royal Society)
and by the University Research  Fund (c/o Victoria University).
Research for ECV is supported by EPEAEK II
in the framework of the grant PYTHAGORAS II - University Research Groups Support
(co-financed $75\%$ by EU funds and $25\%$ by National funds).

\end{document}